\documentclass[final,1p,times]{elsarticle}

\usepackage{times}
\usepackage{color}
\usepackage{float}
\usepackage{amssymb}
\usepackage{epsfig}
\usepackage{mathrsfs}
\usepackage{amsmath}
\usepackage{mathrsfs}
\usepackage{graphicx}
\usepackage{epstopdf}
\usepackage{url}

\usepackage{graphicx}
\usepackage{amsfonts}
\usepackage{epstopdf}

\usepackage{color}
\usepackage{makecell}
\usepackage{array}
\usepackage{booktabs}
\usepackage{tabularx}
\usepackage{multirow}

\usepackage{epsfig}
\usepackage{graphicx}
\usepackage{dcolumn}
\usepackage{bm}
\usepackage{times}
\usepackage{xcolor}
\usepackage{verbatim}
\usepackage{subfigure}
\usepackage{rotate}
\usepackage{float}
\usepackage{xcolor}
\usepackage{lineno,hyperref}
\usepackage{graphics,dcolumn,fleqn,epic}

\usepackage{amssymb}
\usepackage{amsthm}

\newdefinition{defi}{Definition}

\usepackage{soul}

\usepackage{hyperref}
\usepackage{algorithm}
\usepackage{algorithmic}
\biboptions{numbers,sort&compress}

\begin{document}
\begin{frontmatter}


\author[addr1]{Lilei Han}
\author[addr3]{Zhaohua Lin}
\author[addr4]{Qingqing Yin}
\author[addr1,addr2]{Ming Tang\corref{cor1}}
\ead{tangminghan007@gmail.com}
\author[addr1]{Shuguang Guan\corref{cor1}}
\ead{sgguan@phy.ecnu.edu.cn}
\author[addr5,addr6]{Mari\'an Bogu\~n\'a\corref{cor1}}
\ead{marian.boguna@ub.edu}
\cortext[cor1]{Corresponding author}

\address[addr1]{School of Physics and Electronic Science, East China Normal University, Shanghai 200241, China}
\address[addr2]{Shanghai Key Laboratory of Multidimensional Information Processing, East China Normal University, Shanghai 200241, China}
\address[addr3]{Department of Physics, Hong Kong Baptist University, Kowloon Tong, Hong Kong SAR, China}
\address[addr4]{Center for Phononics and Thermal Energy Science, Shanghai Key Laboratory of Special Artificial Microstructure Materials and Technology, School of Physics Science and Engineering, Tongji University, Shanghai 200092, China}
\address[addr5]{Departament de F\'isica de la Mat\`eria Condensada. Mart\'i i Franqu\`es 1, 08028 Barcelona, Spain}
\address[addr6]{Universitat de Barcelona Institute of Complex Systems (UBICS), Universitat de Barcelona, Barcelona, Spain}

\title{Non-Markovian epidemic spreading on temporal  networks}
\author{}

\address{}

\begin{abstract}
Many empirical studies have revealed that the occurrences of contacts associated with human activities are
non-Markovian temporal processes with a heavy tailed inter-event time distribution. Besides, there has been increasing empirical evidence that the infection and recovery rates are time-dependent. However, we lack a comprehensive framework to analyze and understand non-Markovian contact and spreading processes on temporal networks.  In this paper, we propose a general formalism to study non-Markovian dynamics on non-Markovian temporal networks. We find that, under certain conditions, non-Markovian dynamics on temporal networks are equivalent to Markovian dynamics on static networks. Interestingly, this result is independent of the underlying network topology.
\end{abstract}
\end{frontmatter}

\section{Introduction}\label{sec:intro}

Spreading dynamics on complex networks has become a hot research topic during the last decades~\cite{Pastor-Satorras:2015fh}. Due to the covid-19 pandemia, this research field is nowadays in the agenda of public health systems all over the world, as it helps them to take informed decisions on mitigation policies and vaccination campaings. Traditionally, the spreading dynamics is studied with compartmental models such as the susceptible-infected-susceptible (SIS) and susceptible-infected-recovered (SIR) models~\cite{anderson1992infectious}. These models have provided invaluable
insights into the nature of spreading mechanisms, including the diffusion of innovations~\cite{weiss2014adoption,rogers2014diffusion,wang2020diffusion,fibich2016bass}, the spread of cultural fads~\cite{bikhchandani1992theory,banerjee1992simple} and viruses~\cite{dodds2004universal,sheridan2017generalized}. At the same time, network science allows us to understand the interplay between the different spreading mechanisms at play and the underlying network topology, like the absence of epidemic thresholds~\cite{Boguna:2013pd} (and references therein) or the effect of self-similarity~\cite{Serrano:2011lg} and community structure~\cite{salathe2010dynamics}. However, classical models all assume that spreading dynamics are Markovian and take place on static networks. While simplifying the analysis, these assumptions are strongly challenged by empirical observations. On the one hand, a large number of empirical studies have shown that the distributions of infectious and recovery periods of real diseases are far from being exponential, and so Markovian~\cite{bailey1956significance,eichner2003transmission,
nishiura2007infectiousness,chowell2014transmission,
lauer2020incubation,qin2020estimation,mcaloon2020incubation}. On the other hand, a large amount of works show that modeling the underlying substrates where epidemics spread as temporal networks provides a more reasonable representation of real-world complex systems~\cite{holme2012temporal,holme2015modern,lambiotte2016guide}.

In recent years, researchers have made significant strides in addressing the limitations of current models in studying the dynamics of infectious diseases. One approach has been to focus on the effect of non-Markovian dynamics, such as non-Poissonian transmission and recovery processes~\cite{starnini2017equivalence,shkilev2019non,feng2019equivalence}. Theories have been proposed to explain these complex processes and it has been shown that a non-Markovian infection process can dramatically alter the epidemic threshold of the susceptible-infected-susceptible model~\cite{van2013non}. Interestingly, non-Markovian effects in recovery processes can also make the network more resilient against large-scale failures~\cite{lin2020non} and, in some cases, it has been demonstrated that non-Markovian dynamics can be reduced to Markovian dynamics, simplifying the modeling process~\cite{starnini2017equivalence,feng2019equivalence,bottcher2020unifying}. Finally, non-Markovian dynamics may induce an effective complex contagion mechanism, with correlated infectious channels, leading to the appearance of novel exotic epidemic phases~\cite{Hoffmann_2019}.

A parallel line of research aims to understand the effect of temporal networks on spreading dynamics. To better understand complex systems with changing network topologies, various temporal network models have been proposed to replace static, time-aggregated networks~\cite{Perra:2012zk,barrat2013modeling,karsai2014time}. These models reveal that memory effects can raise the epidemic threshold in the SIR model but lower it in the SIS model~\cite{sun2015contrasting}. The most significant factor for spreading dynamics is long-time temporal structures like node and link turnover~\cite{holme2016temporal}. Compared with time-aggregated networks, non-Markovian characteristics in temporal networks can result in a slowdown or acceleration of diffusion~\cite{scholtes2014causality}, it can improve the navigability properties of the system~\cite{Ortiz:2017ho}, or regulate the bursty behavior of dynamic processes~\cite{Garcia-Perez:2015tb}. However, there is currently a lack of research that considers both non-Markovian spreading processes and non-Markovian contact processes simultaneously.

In this paper, we partially fill this gap and consider a non-Markovian SIS dynamics evolving on non-Markovian temporal networks. We develop a mean field approach and show that it can predict well the transient dynamics, the steady state, and the epidemic threshold. Besides, our results show that, in some cases, the steady state of non-Markovian SIS dynamics on temporal networks are equivalent to a Markovian dynamics on the static version of the networks but with an effective infection rate that depends on the details of the particular non-Markovian dynamics. These results provide a deeper understanding of the dynamics of infectious diseases and may help to inform effective disease control and prevention strategies.

\section{Model Description}
\label{sec:model}

To generate a temporal network, we first consider an unweighted and undirected static network as the underlying structure on which temporal interactions take place~\cite{unicomb2021dynamics}. We assume that, in line with the laws of interpersonal networks, node and edge additions and deletions take place on a much longer time scale than the dynamic time scale of events on existing edges. For example, the time scale for making new friends or alienating old friends is typically longer than the time scale for interacting with existing friends. Following this idea, we consider a static network $G(N,E)$ with $N$ nodes and $E$ edges as the underlying structure. However, in many real complex networks, even if the underlying network is static, nodes and edges can be temporarily inactive. In our model, nodes are always active whereas edges obey a stochastic two-state process and can be either active or dormant. Only active edges can be used to propagate the disease from one node to its neighbor. We further consider all edges as identical and statistically independent. The two-state process at each edge is defined by the probability densities $\varphi_{\text{\text{off}}}(\tau)$ and $\varphi_{\text{\text{on}}}(\tau)$, accounting for the random time each edge remains in the dormant (off) or active (on) states, respectively. When the pdf functions $\varphi_{\text{\text{off}}}(\tau)$ and $\varphi_{\text{\text{on}}}(\tau)$ are not exponentials, the system is intrinsically non-Markovian because the instantaneous rate for the transition between the on- and off-states (similarly between the off- and on-states) depends on the time the system has already remained in the on-state (off-state). These rates can be computed as
\begin{equation}
\omega_{\text{on}}(\tau)= \frac{\varphi_{\text{\text{on}}}(\tau)}{\Phi_{\text{on}}(\tau)} \; \; \text{   and   } \; \; \omega_{\text{off}}(\tau)= \frac{\varphi_{\text{\text{off}}}(\tau)}{\Phi_{\text{off}}(\tau)},
\label{eq:rates}
\end{equation}
where $\Phi_{\text{on}}(\tau)$ and $\Phi_{\text{off}}(\tau)$ are the corresponding survival probabilities given by
\begin{equation}
\Phi_{\text{on}}(\tau)=\int_{\tau}^{\infty}
\varphi_{\text{\text{on}}}(\tau')\mathrm{d}\tau' \; \;
\text{   and   } \; \;\Phi_{\text{off}}(\tau)=\int_{\tau}^{\infty}
\varphi_{\text{\text{off}}}(\tau')\mathrm{d}\tau'.
\end{equation}
When on-off dwell times are exponentially distributed with rates $\lambda_{\text{on}}$ and $\lambda_{\text{off}}$, then $\omega_{\text{on}}(\tau)=\lambda_{\text{on}}$ and $\omega_{\text{off}}(\tau)=\lambda_{\text{off}}$ and the process is Markovian. Any other distribution introduces memory in the process. This is, however, the weakest form of memory as it only last between two consecutive events. Yet, this approach is useful in many real-world systems. To guarantee the convergence to the steady state of the process, the pdf functions for the on-off dwell times can take any form as long as their averages $\langle \tau_{\text{on}} \rangle$ and $\langle \tau_{\text{off}} \rangle$ are finite.

In this paper, we are interested in the temporal evolution of an epidemic outbreak starting from a small fraction of the population infected at $t=0$. However, we assume that, prior to the start of the outbreak, the on-off dynamics of the network is already at its steady state. In this condition, the probability to find any given edge in the on-state is~\cite{BOGUNA2000475}
\begin{equation}
P_{\text{on}}=\frac{\langle \tau_{\text{on}} \rangle}{\langle \tau_{\text{on}} \rangle+\langle \tau_{\text{off}} \rangle},
\end{equation}
and, given that we find the edge in the on-state, the pdf of the time since the edge entered the on-state is~\cite{BOGUNA2000475}
\begin{equation}
\psi_{\text{on}}(\tau)=\frac{\Phi_{\text{on}}(\tau)}{\langle \tau_{\text{on}} \rangle}.
\end{equation}
Finally, it is important to stress that the two-state dynamics defining the temporal network is completely blind to the epidemic state of the system, whereas, as discussed later, the oposite is not true.

On top of the temporal network described above, we implement the non-Markovian SIS model. In this model, infected nodes recovers spontaneously after a random time $\tau$ that obeys the probability density function $\varphi_{\text{rec}}(\tau)$. Infection events, on the other hand, can only take place through active edges when one of the nodes is in the infected state and the other in the susceptible state, defining an active infectious edge (AIE). Once the infection event is initiated, the actual infection of the susceptible node takes a random time governed by the pdf $\varphi_{\text{inf}}(\tau)$. Similarly to the on-off dynamics, we can define the instantaneous recovery and infection rates $\omega_{\text{rec}}(\tau)$ and $\omega_{\text{inf}}(\tau)$ as in Eqs~\eqref{eq:rates}. In this work, as opposed to complex contagion mechanisms, we consider a simple infection scheme where the total infection rate of a node is the sum of the infection rates of incident edges.

\begin{figure}[t]
\centering
\includegraphics[width=1\textwidth]{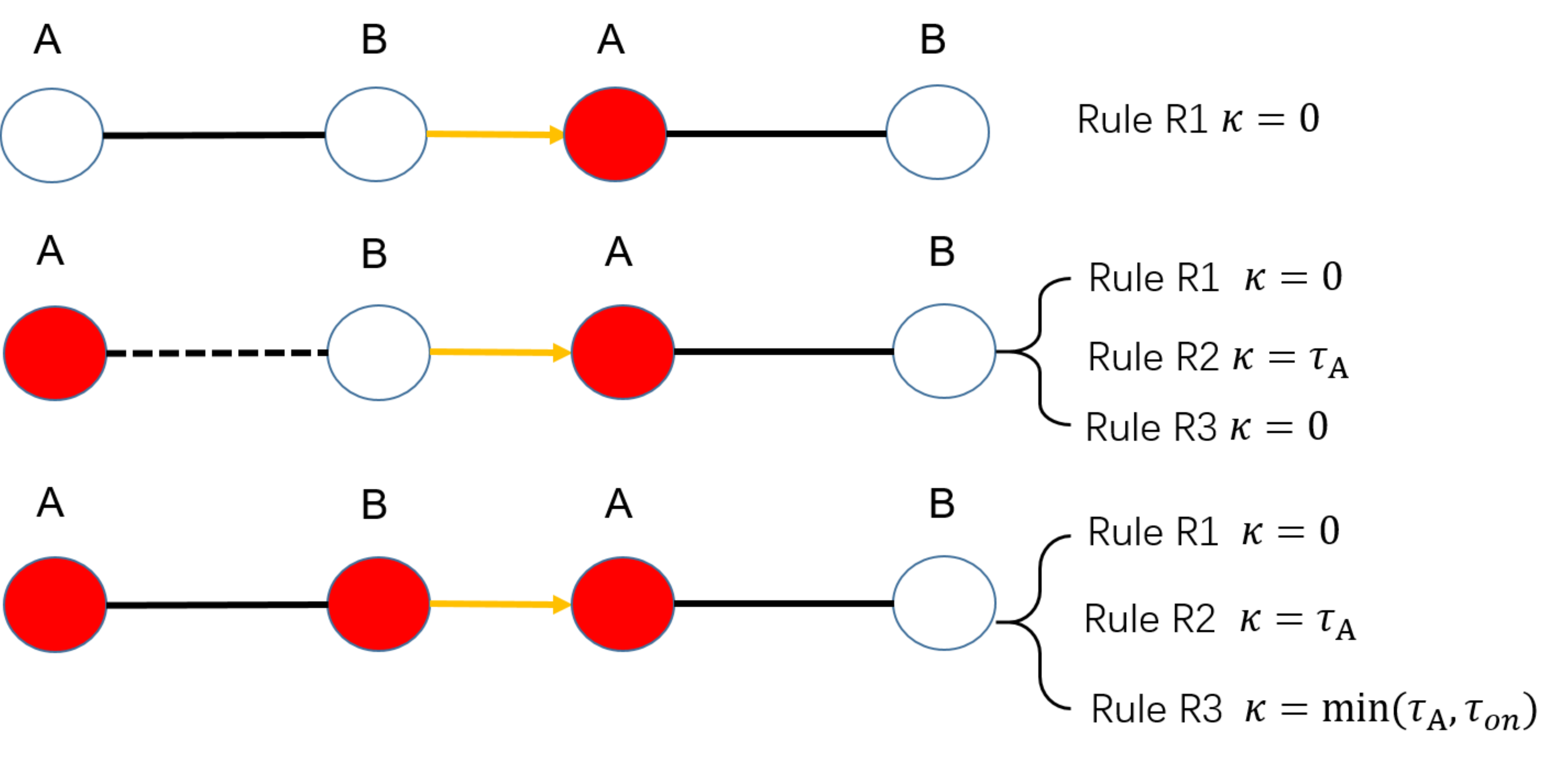}
\renewcommand{\figurename}{FIG.}
\caption{Three possible ways to generate a new AIE. Red and white circles
represent infected and susceptible nodes, respectively, whereas solid and
dashed lines indicate active and dormant edges, respectively. We also show
how to define the onset of the infection from A to B for the three rules
used in this paper. Rule 1 sets this time always to zero. Rule 2 assigns to
the newborn AIE the infected dwell time of node A. In Rule 3, the onset of
the infection is set as the minimum between the infected dwell time of node
A and the active dwell time of the edge connecting nodes A and B.}
\label{fig:rule}
\end{figure}

When temporal events are not exponentially distributed, it is important to precisely define when each event is initiated. Recovery events of infected nodes do not depend on the network connectivity and, thus, are initiated at the very moment nodes get infected. This is not the case for infection events, for which it is not clear where to set the onset of the infection. One can adopt a node-centric approach from the point of view of the susceptible node and set the onset time at the moment an AIE is created. We call this approach ``Rule 1'' (R1)~\cite{Boguna:2014tk}. Figure~\ref{fig:rule} shows three examples of the application of Rule 1: in the first one, both nodes A and B are originally susceptible and the edge is active. Then individual A becomes infected by one of his neighbors other than B, generating a new AIE and defining this time as the onset of the infection. In the second example, node A is infected, B is susceptible and the edge connecting them is dormant. Then, the edge becomes active while node A is still infected, generating an AIE and setting at this moment the onset of the infection. In the third example, both nodes A and B are infected and the edge is active. Then node B recovers and becomes susceptible again. This again defines a new AIE and sets the onset of the infection from A to B. Unlike Rule 1, we can adopt a node-centric approach from the point of view of the infected node and set the onset of the infection at the time the node gets infected, regardless of the state of its edges or neighbors. We call this approach ``Rule 2'' (R2)~\cite{Boguna:2014tk}. The approach takes into account the different infectivity phases an infected individual can go through. For instance, for diseases with incubation periods, when the individual is infected but not very infectious. Finally, we can adopt a mixed approach, node-centered at the infected node but modulated by the activity of edges. We call this approach ``Rule 3'' (R3). In this case, we set the onset of the infection at the time the infected node makes contact to its neighbors, regardless of their state. So R3 depends both on the state of the infected node itself as well as on the state of its edges. Figure~\ref{fig:rule} shows also the onset of the infection for rules R2 and R3 in the different cases. Other types of approaches can induce an effective complex contagion and the split of the epidemic phase transition into new intermediate phases~\cite{Boguna:2014tk}

\section{Non-Markovian Mean-field approach}
\label{sec:theory}

Here, we extend the mean field approximation developed in
Ref.\cite{feng2019equivalence} for static networks to
the case of the temporal networks described in the previous section.
Due to the non-Markovia dynamics, to describe the temporal evolution of
the epidemic we have to keep track of the dwell time of nodes in each state,
infected or susceptible. We then define $I_i(\tau;t)d\tau$ as the
probability to find node $i$ infected at time $t$ and that, simultaneously,
the time since it became infected is within the interval $(\tau,\tau+d\tau)$.
Similarly, we define $S_i(\tau;t)$ as the probability to find node $i$
susceptible at time $t$ and that, simultaneously, the time since it became
susceptible is within the interval $(\tau,\tau+d\tau)$. Notice that
functions $I_i(\tau;t)$ and $S_i(\tau;t)$ are probability densities with
respect to the variable $\tau$ but not $t$, so that the prevalence of node
$i$ at time $t$, $\rho_i(t)$, defined as the probability to find node $i$
infected at time $t>0$ is
\begin{equation}
\rho_i(t)=\int_0^t I_i(\tau;t)d\tau,
\label{prevalence0}
\end{equation}
and the global prevalence $\rho(t)=N^{-1} \sum_{i=1}^N \rho_i(t)$. Hereafter, as initial conditions, we assume that node $i$ is infected at time $t=0$ with probability $\rho_i(t=0)=\rho_{i,0}$.

Assuming that the infections from different edges are statistically
independent events and working at the mean field level, in
Ref.\cite{feng2019equivalence} it was shown that functions
$I_i(\tau;t)$ and $S_i(\tau;t)$ satisfy the partial differential equations
\begin{equation}
\left(\frac{\partial}{\partial\tau}+\frac{\partial}{\partial t}\right)I_i(\tau;t)=-\omega_{\text{rec}}(\tau) I_i(\tau;t)
\label{partial1}
\end{equation}
and
\begin{equation}
\left(\frac{\partial}{\partial\tau}+\frac{\partial}{\partial t}\right)S_i(\tau;t)=- \sum_{j=1}^N a_{ij} \phi_{i\leftarrow j}(\tau;t) S_i(\tau;t)
\label{partial2}
\end{equation}
where $\phi_{i\leftarrow j}(\tau;t)$ is the instantaneous infection rate
from node $j$ to $i$, which depend on the specific rule used and the on-off
dynamics on the edges. Equations~\eqref{partial1} and~\eqref{partial2} are
supplemented by the boundary conditions
\begin{equation}
S_i(0;t)=\int_0^t \omega_{\text{rec}}(\tau) I_i(\tau;t)d\tau
\label{boundary1}
\end{equation}
and
\begin{equation}
I_i(0;t)=\sum_{j=1}^N a_{ij}\int_0^t S_i(\tau;t)  \phi_{i\leftarrow j}(\tau;t) d\tau,
\label{boundary2}
\end{equation}
Notice that these boundary conditions are slightly different from the ones
in Ref.\cite{feng2019equivalence} because we consider that
dwell times are bounded in the interval $\tau \in(0,t)$.

\begin{figure}
\centering
\includegraphics[width=1\textwidth]{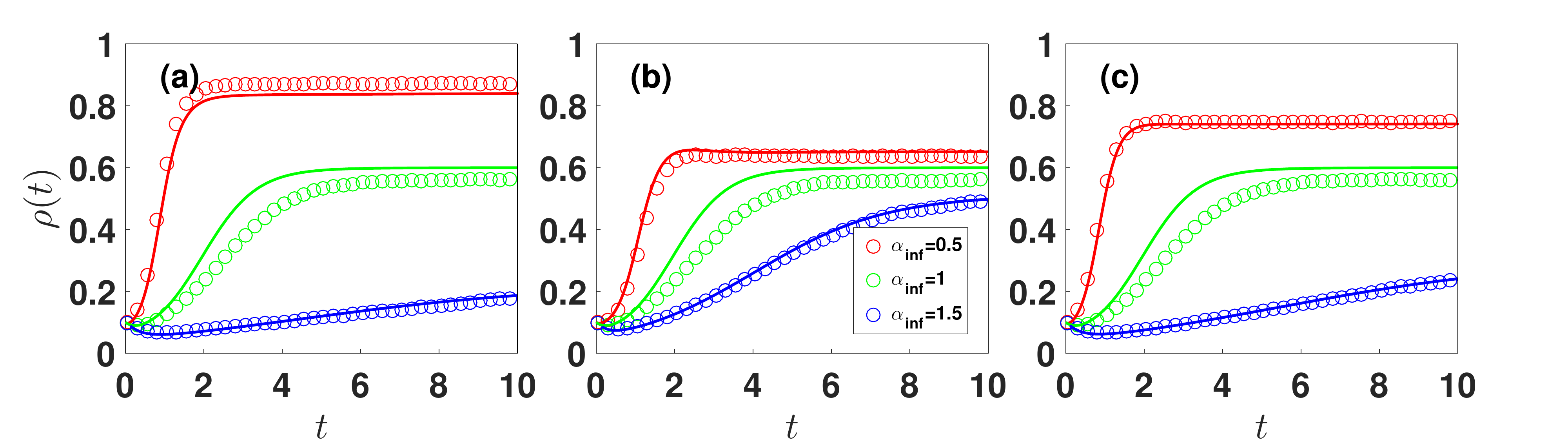}
\renewcommand{\figurename}{FIG.}
\caption{Comparison between simulation and theoretical results of the
temporal evolution of the density of infected nodes $\rho(t)$ on a
random-regular network with $N=10000$ and $\langle k \rangle=10$ for
rules R1 (a), R2 (b), and R3 (c). Red circles, green squares, and blue
triangles represent the simulation results for $\alpha_{\text{inf}}=0.5,1,$
and $1.5$, respectively and solid lines are theoretical results.
Other parameters are $\langle \tau_{\text{off}}\rangle=\langle\tau_{\text{on}}\rangle=1,
\alpha_{\text{off}}= \alpha_{\text{on}}=5$, $u=0.5$ and $\delta=1$.}
\label{fig:figure3}
\end{figure}

Finally, to close these equations, we need to find an expression for the
instantaneous infection rate from node $j$ to susceptible node $i$
(with susceptible dwell time $\tau$) at time $t$,
$\phi_{i\leftarrow j}(\tau;t)$. This rate depends on the particular rule used
to determine the onset of the infection event. For R1, this time is the
smallest between the susceptible dwell time of node $i$, the infected dwell
time of node $j$, and the active dwell time of the edge connecting nodes $i$
and $j$. Thus, we can write
\begin{equation}
\phi_{i\leftarrow j}^{\text{R1}}(\tau;t)=P_{\text{on}}\int_{0}^\infty d \tau' \frac{\Phi_{\text{on}}(\tau')}{\langle \tau_{\text{on}} \rangle}
\int_0^t d\tau'' I_j(\tau'';t) \omega_{\text{inf}}\left(\min{(\tau,\tau',\tau'')}\right)
\label{rateR1}
\end{equation}
For R2, the onset of the infection event is determined by the dwell time in
the infected state of node $j$ at time $t$. Therefore, we write
\begin{equation}
\phi_{i\leftarrow j}^{\text{R2}}(\tau;t)=\phi_{i\leftarrow j}^{\text{R2}}(t)=P_{\text{on}}
\int_0^t d\tau'' I_j(\tau'';t) \omega_{\text{inf}}\left(\tau'' \right).
\label{rateR2}
\end{equation}
For R3, this time is the smaller between the infected dwell time of node $j$,
and the active dwell time of the edge connecting nodes $i$ and $j$. Thus,
we have
\begin{equation}
\phi_{i\leftarrow j}^{\text{R3}}(\tau;t)=\phi_{i\leftarrow j}^{\text{R3}}(t)=P_{\text{on}}\int_{0}^\infty d \tau' \frac{\Phi_{\text{on}}(\tau')}{\langle \tau_{\text{on}} \rangle}
\int_0^t d\tau'' I_j(\tau'';t) \omega_{\text{inf}}\left(\min{(\tau',\tau'')}\right)
\label{rateR3}
\end{equation}
The set of equations Eqs.~\eqref{prevalence0}-~\eqref{rateR3} form a closed
set of equations that enable to find the temporal evolution of the
prevalence $\rho_i(t)$ at the mean field level, as shown in
Fig.~\ref{fig:figure3}. From this analysis, we already conclude that
the only influence of the off-distribution $\varphi_{\text{\text{off}}}(\tau)$
on the SIS dynamics is through its average value, which affect the
probability to find an edge in the on-state $P_{\text{on}}$.

It is illustrative to see how this formalism recovers the Markovian SIS model
at the mean field level for static networks. In that case,
$\omega_{\text{rec}}(\tau)=\delta$, $\omega_{\text{inf}}(\tau)=\lambda$ and
$P_{\text{on}}=1$. Integrating Eq.~\eqref{partial1} with respect to
$\tau \in (0,t)$ leads to the differential equation
\begin{equation}
\frac{d \rho_i(t)}{dt}=-\delta \rho_i(t)+I_i(0;t)=-\delta \rho_i(t)+\sum_{j=1}^N a_{ij}\int_0^t S_i(\tau;t)  \phi_{i\leftarrow j}(\tau;t) d\tau
\label{prevalence1}
\end{equation}
In the Markovian case, the instantaneous infection rate from node $j$ to
$i$ is independent of the rule used and takes the simple form
\begin{equation}
\phi_{i\leftarrow j}^{\text{R1}}(\tau;t)=\phi_{i\leftarrow j}^{\text{R2}}(\tau;t)=\phi_{i\leftarrow j}^{\text{R3}}(\tau;t)=\lambda \rho_j(t).
\end{equation}
Using this result in Eq.~\eqref{prevalence1} we obtain
\begin{equation}
\frac{d \rho_i(t)}{dt}=-\delta \rho_i(t)+\lambda \sum_{j=1}^N a_{ij} [1-\rho_i(t)]\rho_j(t),
\end{equation}
which is the mean field approximation of the SIS model developed in
Ref.~\cite{van2008virus}. At the steady state, the prevalence
satisfy
\begin{equation}
\rho_i^{\text{st}}=\lambda_{\text{eff}} \sum_{j=1}^N a_{ij} [1-\rho_i^{\text{st}}]\rho_j^{\text{st}}.
\label{steadyMarkovian}
\end{equation}
with $\lambda_{\text{eff}}=\lambda/\delta$.

\subsection{The steady state solution}

At the steady state, functions $I_i(\tau;t)$, $S_i(\tau;t)$, and
$\phi_{i\leftarrow j}(\tau;t)$ become independent of $t$ and we can find the
following explicit expressions from Eqs.~\eqref{partial1} and~\eqref{partial2}
\begin{equation}
I_i^{\text{st}}(\tau)=\frac{\rho_i^{\text{st}} }{\langle \tau_{\text{rec}}\rangle} \Phi_{\text{\text{rec}}}(\tau) \; \; \text{and} \; \;
S_i^{\text{st}}(\tau)=\frac{\rho_i^{\text{st}}}{\langle \tau_{\text{rec}}\rangle} \exp{\left( -\sum_j a_{ij} \int_0^\tau \phi_{i\leftarrow j}^{\text{st}}(\tau') d\tau' \right)}.
\label{eq:steady}
\end{equation}
Integrating the last equation we obtain
\begin{equation}
1-\rho_i^{\text{st}}=\frac{\rho_i^{\text{st}}}{\langle \tau_{\text{rec}}
\rangle} \int_0^{\infty}  \exp{\left( -\sum_j a_{ij} \int_0^\tau
\phi_{i\leftarrow j}^{\text{st}}(\tau') d\tau' \right)}d\tau.
\label{prevalence_closed}
\end{equation}
Notice that combining the expression for $I_i^{\text{st}}(\tau)$ in Eq.~\eqref{eq:steady} and the general expression for $\phi_{i\leftarrow j}(\tau;t)$, we can write that, at the steady state, $\phi_{i\leftarrow j}^{\text{st}}(\tau) = \frac{\rho_j^{\text{st}}
P_{\text{on}}}{\langle \tau_{\text{rec}}\rangle} f(\tau)$, where
$f(\tau)$ is a function that depends on the particular rule used but
does not depend on the states of nodes $i$ and $j$. Thus, Eqs.~\eqref{prevalence_closed}
form set of closed equations for $\rho_i^{\text{st}}$ $\forall i=1,\cdots,N$.

To go further, we need to specify the particular rule used. In the case of R2,
function $f(\tau)$ is independent of $\tau$, taking the value
\begin{equation}
f(\tau)=\int_0^{\infty} d\tau'' \Phi_\text{rec}(\tau'')
\omega_{\text{inf}}(\tau'').
\label{steadyR2_f}
\end{equation}
Using this result, Eq.~\eqref{prevalence_closed} reduces to
\begin{equation}
\rho_i^{\text{st}}=\lambda_{\text{eff}}^{R2} \sum_{j=1}^N a_{ij} [1-\rho_i^{\text{st}}]\rho_j^{\text{st}},
\label{steadyR2}
\end{equation}
with
\begin{equation}
\lambda_{\text{eff}}^{R2}=P_{\text{on}} \int_0^\infty \Phi_{\text{\text{rec}}}(\tau) \frac{\varphi_{\text{\text{inf}}}(\tau)}{\Phi_{\text{\text{inf}}}(\tau)} d\tau.
\label{lambdaeffR2}
\end{equation}
Equation~\eqref{steadyR2} is identical to the mean field steady state of the
non-Markovian SIS model in Eq.~\eqref{steadyMarkovian} but with an effective
infection rate given by $\lambda_{\text{eff}}^{R2}$ so that, at the steady
state, we can replace a non-Markovian SIS epidemic process on a temporal
network by a Markovian one on a static network. It is also interesting to
notice that, for rule R2, the effect of the temporal nature of the network
has only a minor effect on the dynamics through the probability
$P_{\text{on}}$. However, the particular details of the on-off dynamics are
very relevant to determine the time needed to reach the steady state, or the
time window one must observe the system to make sure that the steady state
is sufficiently sampled.

In the case of R3, function $f(\tau)$ is also independent of $\tau$ and takes the form
\begin{equation}
f(\tau)=
\int_{0}^\infty d \tau' \frac{\Phi_{\text{on}}(\tau')}
{\langle \tau_{\text{on}} \rangle}
\int_0^{\infty} d\tau'' \Phi_\text{rec}(\tau'')
\omega_{\text{inf}}\left(\min{(\tau',\tau'')}\right),
\label{steadyR3_f}
\end{equation}
so that Eq.~\eqref{prevalence_closed} reduces to
\begin{equation}
\rho_i^{\text{st}}=\lambda_{\text{eff}}^{R3} \sum_{j=1}^N a_{ij} [1-\rho_i^{\text{st}}]\rho_j^{\text{st}},
\label{steadyR3}
\end{equation}
with
\begin{equation}
\lambda_{\text{eff}}^{R3}=P_{\text{on}}
\int_0^\infty
\frac{\Phi_{\text{on}}(\tau')}{\langle \tau_{\text{on}} \rangle} d\tau'
\int_0^\infty \Phi_{\text{rec}}(\tau'')
\omega_{\text{inf}}(\min(\tau',\tau''))d\tau''.
\label{lambdaeffR3}
\end{equation}
As in the previous case, the steady state of the non-Markovian SIS dynamics
under R3 can be reduced to the Markovian case with the effective infection
rate $\lambda_{\text{eff}}^{R3}$. Unlike R2, here the on-off dynamics of
edges plays a more relevant role because the distribution of dwell times of
edged in the on state has a strong influence on the definition of the
effective rate $\lambda_{\text{eff}}^{R3}$.

In the case of rule R1, function $f(\tau)$ is not constant and it is not
possible to reduce the steady state to the non-Markovian case. Besides, in
this case, the assumption about the independence between different
infectious channels does not hold. We can understand this phenomenon with a
simple example: consider an infected node A with three infected neighbors B,
C, and D, connected by active edges. If node A recovers, then the infections
from nodes B, C, and D to now susceptible node A will start simultaneously
and, therefore, become strongly correlated. However, it is still possible to
find an effective infection rate that becomes exact near the phase
transition. Indeed, in this case the density of infected nodes is very low
and the number of infected neighbors is small so that the correlations
between different edges become irrelevant. We can then use the formalism
in Ref.~\cite{starnini2017equivalence}, where the effective
infection rate from node $j$ to $i$ is defined as
\begin{equation}
\lambda_{\text{eff}}=P_{\text{on}} \langle \tau_{\text{rec}}\rangle \langle \omega_{\text{inf}}(\tau) | \text{node $i$ susceptible, $j$ infected, edge $ij$ active} \rangle,
\end{equation}
where $\tau$ is the time since the infection process started given that
node $i$ is susceptible, $j$ infected, and the edge between them active.
For rule R1 at the steady state, the only information about this time is
that it must be shorter than the time $j$ takes to infect $i$, also shorter
than the time $j$ takes to recover, and shorter than the time edge $ij$
remains in the active state. Therefore, the pdf of this time is given by
\begin{equation}
\psi(\tau | \text{node $i$ susceptible, $j$ infected, edge $ij$ active})= \frac{\Phi_{\text{inf}}(\tau) \Phi_{\text{rec}}(\tau) \Phi_{\text{on}}(\tau)}{\int_0^\infty \Phi_{\text{inf}}(\tau) \Phi_{\text{rec}}(\tau) \Phi_{\text{on}}(\tau) d\tau}
\end{equation}
and the effective infection rate for rule R1 becomes
\begin{equation}
\lambda_{\text{eff}}^{R1}=P_{\text{on}} \langle \tau_{\text{rec}}\rangle \frac{\int_0^\infty \varphi_{\text{inf}}(\tau) \Phi_{\text{rec}}(\tau) \Phi_{\text{on}}(\tau) d\tau}{\int_0^\infty \Phi_{\text{inf}}(\tau) \Phi_{\text{rec}}(\tau) \Phi_{\text{on}}(\tau) d\tau}.
\label{lambdaeffR1}
\end{equation}
We then expect that near the critical point, the behavior of the
non-Markovian SIS on a temporal network is equivalent to the Markovian
SIS on a static network with the effective infection rate
$\lambda_{\text{eff}}^{R1}$. Then, we can use the expression in
Eq.~\eqref{lambdaeffR1} to find the exact position of the critical point in
terms of the critical point of the non-Markovian SIS on the static version
of the network.

\begin{figure}[t]
\centering
\includegraphics[width=1\textwidth]{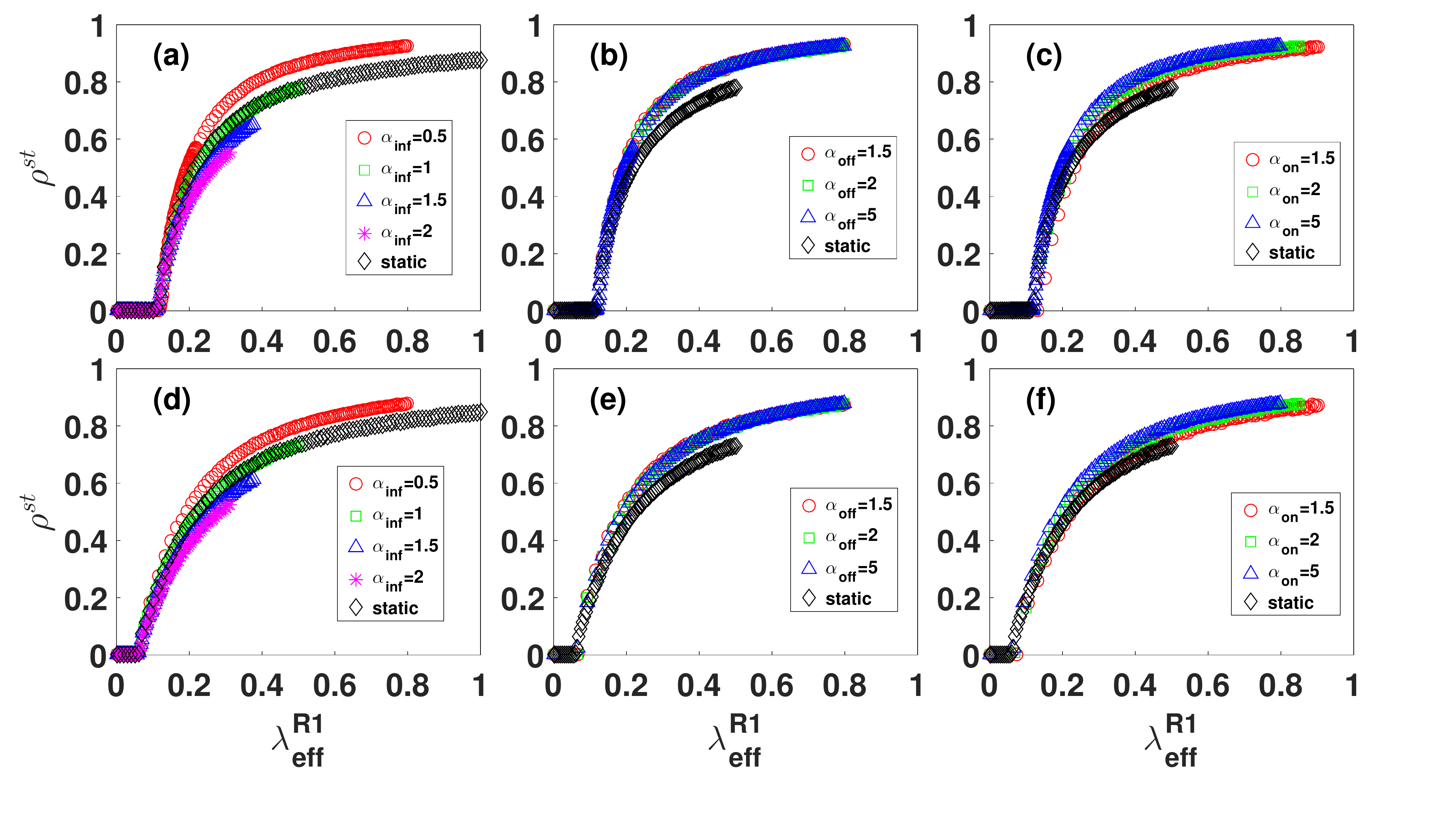}
\renewcommand{\figurename}{FIG.}
\caption{Steady-state prevalence $\rho^{\text{st}}$ under rule R1 as a
function of $\lambda_{\text{eff}}^{R1}$ in Eq.~\eqref{lambdaeffR1}
for different values of the parameters
$\alpha_{\text{inf}}$, $\alpha_{\text{off}}$ and $\alpha_{\text{on}}$.
Panels (a), (b), and (c) show results for Erd{\H{o}}s-R{\'{e}}nyi networks
with $\langle k\rangle=10$ and size $N=10000$.  Panels (d), (e), and (f)
show results of Barab{\'{a}}si-Albert networks with $N=10000$.
In panels (a) and (d) we fix $\alpha_{\text{off}}=5$, $\alpha_{\text{on}}=5$
 and change $\alpha_{\text{inf}}$.
In panels (b) and (e) we fix $\alpha_{\text{inf}}=0.5$,
$\alpha_{\text{on}}=5$ and change $\alpha_{\text{off}}$.
Finally, in panels (c) and (f) we fix $\alpha_{\text{inf}}=0.5$,
$\alpha_{\text{off}}=5$ and change $\alpha_{\text{on}}$.
In all cases, we set $\langle \tau_{\text{off}} \rangle=\langle
\tau_{\text{on}} \rangle=1$, $u=0.5$, and $\delta=1$. Black empty diamonds
correspond to the Markovian SIS dynamic on the static version of the network
with the infection rate $\lambda_{\text{eff}}^{R1}$.}
\label{fig:figure4}
\end{figure}
\section{Simulation Results}
\label{sec:sim}

We check the validity of the theory developed in the previous section by
means of extensive numerical simulations of the non-Markovian SIS dynamics
on temporal networks. As suggested by empirical observations, we consider both the dormant and active
times of edges to be power law distributed. Specifically, we use the Lomax
distribution ~\cite{mancastroppa2019burstiness} as
\begin{equation}
\label{on_off}
\varphi(\tau)=\frac{\alpha \xi^{\alpha}}{(\tau+\xi)^{(1+\alpha)}},
\end{equation}
with exponent $\alpha>1$ and $\xi=(\alpha-1) \langle \tau \rangle$. Infections are modeled with a Weibull distribution with shape parameter $\alpha_{\text{inf}}$ and scale parameter $u$.
\begin{equation}
\varphi_{\text{inf}}(\tau)=\alpha_\text{inf} u^{\alpha_\text{inf}}\tau^{\alpha_\text{inf}-1}e^{-(u \tau)^{\alpha_\text{inf}}}.
\end{equation}
This is a useful choice that interpolates between a strongly picked
distribution when $\alpha_{\text{inf}}>1$ to a long tailed distribution
when $\alpha_{\text{inf}}<1$, recovering the Markovian case when
$\alpha_{\text{inf}}=1$. Finally, we consider a Markovian recovery process
with rate $\delta$, $\varphi_{\text{rec}}(\tau)=\delta e^{-\delta \tau}$.

We first check the ability of our formalism to describe the short term
temporal evolution of the dynamics. Specifically, we perform numerical
simulations initially infecting $10\%$ of nodes in the network.
Figure~\ref{fig:figure3} shows results from numerical simulations for the
three rules compared to the numerical solution of
Eqs.~\eqref{prevalence0}-~\eqref{rateR3}. As it can be observed,
the non-Markovian mean field approach agrees very well with numerical
simulations. In particular, it is able to predict the characteristic time
scale to reach the steady state and a non trivial non-monotonous behavior
when the probability density of infection times is picked around it average
(so that $\alpha_\text{inf}>1$).
\begin{figure}[t]
\centering
\includegraphics[width=1\textwidth]{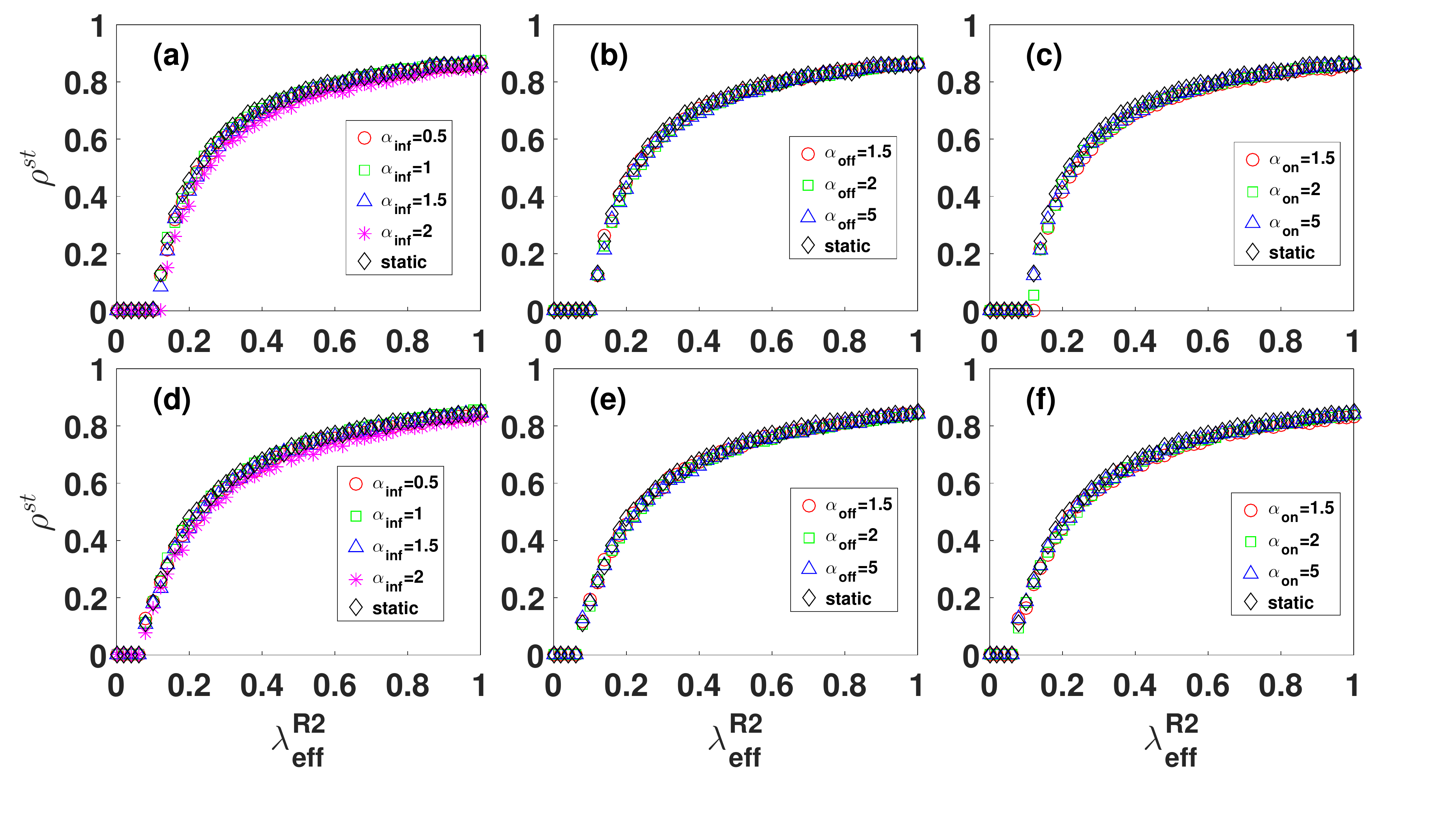}
\renewcommand{\figurename}{FIG.}
\caption{Steady-state prevalence $\rho^{\text{st}}$ under rule R2 as a
function of $\lambda_{\text{eff}}^{R2}$ in Eq.~\eqref{lambdaeffR2}) for different values of the parameters
$\alpha_{\text{inf}}$, $\alpha_{\text{off}}$ and $\alpha_{\text{on}}$. Panels (a), (b), and (c) show results for Erd{\H{o}}s-R{\'{e}}nyi networks with $\langle k\rangle=10$ and size $N=10000$.  Panels (d), (e), and (f) show results of Barab{\'{a}}si-Albert networks with $N=10000$.
In panels (a) and (d) we fix $\alpha_{\text{off}}=5$, $\alpha_{\text{on}}=5$
and change $\alpha_{\text{inf}}$.
In panels (b) and (e) we fix $\alpha_{\text{inf}}=0.5$, $\alpha_{\text{on}}=5$
 and change $\alpha_{\text{off}}$.
Finally, in panels (c) and (f) we fix $\alpha_{\text{inf}}=0.5$,
$\alpha_{\text{off}}=5$ and change $\alpha_{\text{on}}$.
In all cases, we set $\langle \tau_{\text{off}} \rangle=\langle
\tau_{\text{on}} \rangle=1$, $u=0.5$, and $\delta=1$. Black empty diamonds
correspond to the Markovian SIS dynamic on the static version of the network
with the infection rate $\lambda_{\text{eff}}^{R2}$.}
\label{fig:figure5}
\end{figure}

\begin{figure}[t]
\centering
\includegraphics[width=1\textwidth]{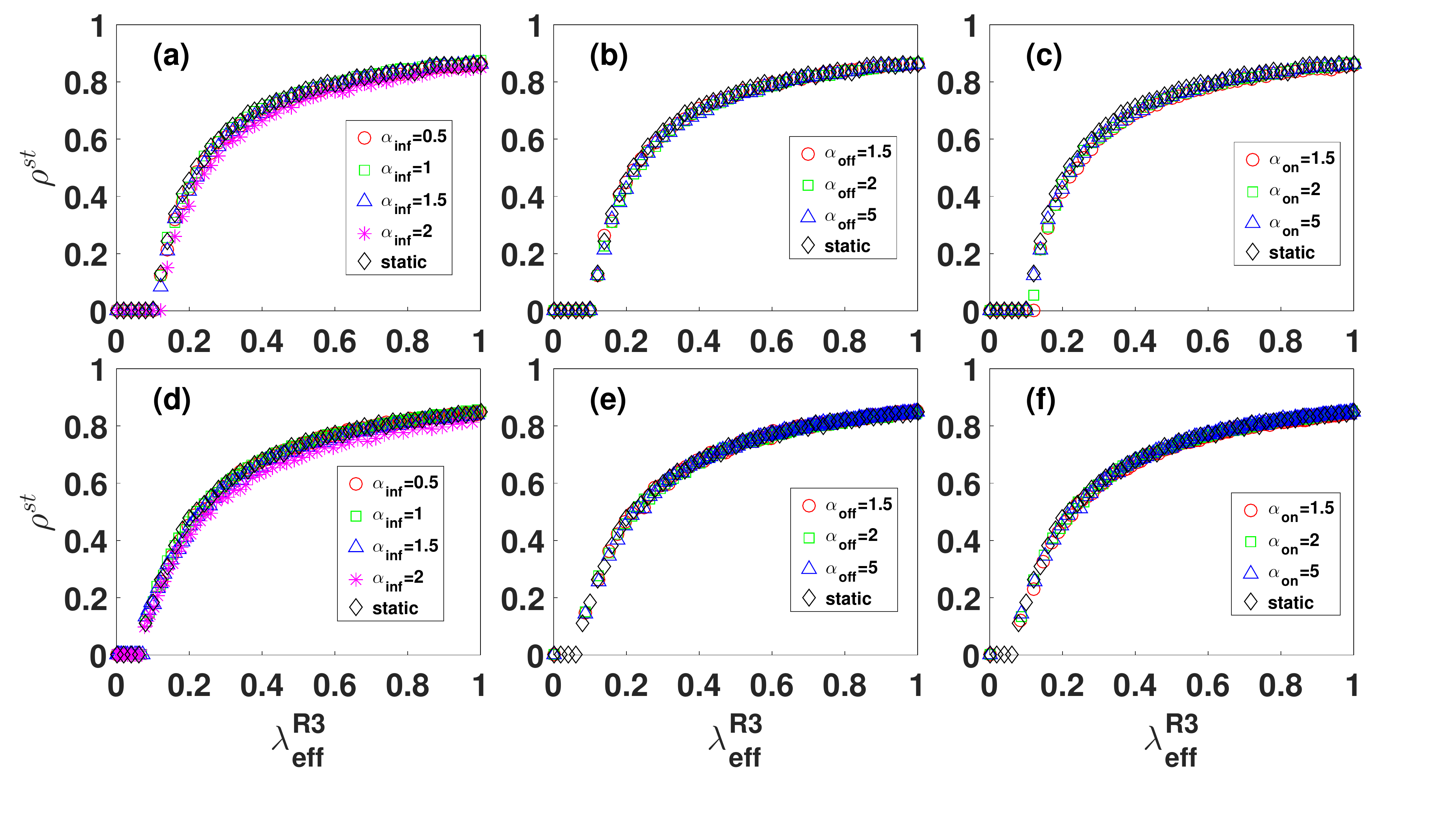}
\renewcommand{\figurename}{FIG.}
\caption{Steady-state prevalence $\rho^{\text{st}}$ under rule R3 as a
function of $\lambda_{\text{eff}}^{R3}$ in Eq.~\eqref{lambdaeffR3})
for different values of the parameters $\alpha_{\text{inf}}$, $\alpha_{\text{off}}$ and $\alpha_{\text{on}}$. Panels (a), (b), and (c) show results for Erd{\H{o}}s-R{\'{e}}nyi networks with $\langle k\rangle=10$ and size $N=10000$.  Panels (d), (e), and (f) show results of Barab{\'{a}}si-Albert networks with $N=10000$.
In panels (a) and (d) we fix $\alpha_{\text{off}}=5$, $\alpha_{\text{on}}=5$ and change $\alpha_{\text{inf}}$.
In panels (b) and (e) we fix $\alpha_{\text{inf}}=0.5$, $\alpha_{\text{on}}=5$ and change $\alpha_{\text{off}}$.
Finally, in panels (c) and (f) we fix $\alpha_{\text{inf}}=0.5$,
$\alpha_{\text{off}}=5$ and change $\alpha_{\text{on}}$.
In all cases, we set $\langle \tau_{\text{off}} \rangle=\langle
\tau_{\text{on}} \rangle=1$, $u=0.5$, and $\delta=1$. Black empty diamonds
correspond to the Markovian SIS dynamic on the static version of the network
with the infection rate $\lambda_{\text{eff}}^{R3}$.}
\label{fig:figure6}
\end{figure}

Next, we study the steady state of the dynamics and compare numerical
simulations with different values of the parameters of the SIS dynamics and
the temporal network. Figure~\ref{fig:figure4} shows the steady prevalence
as a function of the effective infection rate for rule R1,
$\lambda_{\text{eff}}^{R1}$. As discussed above, the prevalence for different
values of the parameters do not colapse into a single curve, which indicates
that the non-Markovian SIS model cannot be reduced to the Markovian case
under rule R1. However, all curves do colapse nicely when the prevalence
approaches zero. This means that, as predicted, the effective infection
rate $\lambda_{\text{eff}}^{R1}$ is able to recover the exact value of the
epidemic threshold in terms of the epidemic threshold of the Markovian dynamics
on the static version of the network. The situation is different in the case
of rules R2 and R3. In both cases, our theory predicts that the non-Markovian
dynamics can be reduced to the Markovian one on a static network with
effective infection rates $\lambda_{\text{eff}}^{R2}$ and
$\lambda_{\text{eff}}^{R3}$. Figures~\ref{fig:figure5} and~\ref{fig:figure6}
show numerical simulations for rules R2 and R3 where the colapse of the
different prevalence curves is evident, thus corroborating our predictions.

\section {Conclusion and Discussion}
\label{sec:con}

Empirical evidence demonstrates that non-Markovian dynamics and temporal
networks are prevalent in real complex systems. However, these properties,
particularly those associated with memory effects, are frequently overlooked
in scientific literature. This is partially due to the analytical and
computational challenges that these properties entail, but the major
obstacle when dealing with non-Markovian dynamics is our lack of knowledge
about how memory is implemented. In this study, we explored three different
possibilities, but numerous other possibilities exist, and it is challenging
to determine which ones apply to actual systems. This has significant
consequences because, as our work has shown, the specifics of the rule
utilized dictate the fate of the dynamics. We have demonstrated that some of
these rules are ``simple'' in that an effective parameter can encode the
non-Markovian and temporal properties of the dynamics. However, in other
cases, this is only possible near the critical point, allowing us to recover
the critical threshold of the dynamics. In light of these findings, it is
uncertain whether the concept of universality class can be extended to
general non-Markovian dynamics.

\section{Acknowledgments} \label{sec:intro}
This work was supported by the National ten thousand talents plan youth top
talent project, the National Natural Science Foundation of
China (Grant Nos. 11975099, 12231012, 11875132),  the Science and Technology Commission
 of Shanghai Municipality (Grant No.22DZ2229004) and the China Scholarship
 Council (Grant No. 202006140147). M.~B. acknowledge support from:
 Grant TED2021-129791B-I00 funded by MCIN/AEI/10.13039/501100011033 and
 the ``European Union NextGenerationEU/PRTR'''; Grant PID2019-106290GB-C22
 funded by\\ MCIN/AEI/10.13039/501100011033; Generalitat de
 Catalunya grant number 2021SGR00856 and the ICREA Academia award, funded
 by the Generalitat de Catalunya.

\section*{References}

\end{document}